# Tailoring Lattice Strain and Ferroelectric Polarization of Epitaxial BaTiO$_3$ Thin Films on Si(001)

Jike Lyu, Ignasi Fina, Raul Solanas, Josep Fontcuberta & Florencio Sánchez*

Institut de Ciència de Materials de Barcelona (ICMAB-CSIC), Campus UAB, Bellaterra 08193, Barcelona, Spain

* fsanchez@icmab.es

Ferroelectric BaTiO$_3$ films with large polarization have been integrated with Si(001) by pulsed laser deposition. High quality *c*-oriented epitaxial films are obtained in a substrate temperature range of about 300 °C wide. The deposition temperature critically affects the growth kinetics and thermodynamics balance, resulting on a high impact in the strain of the BaTiO$_3$ polar axis, which can exceed 2% in films thicker than 100 nm. The ferroelectric polarization scales with the strain and therefore deposition temperature can be used as an efficient tool to tailor ferroelectric polarization. The developed strategy overcomes the main limitations of the conventional strain engineering methodologies based on substrate selection: it can be applied to films on specific substrates including Si(001) and perovskites, and it is not restricted to ultrathin films.

Lattice strain generally causes dramatic effects on the properties of ferroelectrics. In particular, epitaxial strain has permitted a notable enhancement of the ferroelectric polarization and the Curie temperature.[1,2] However, conventional substrate-based strain engineering is restricted to relatively small ranges of strain and film thickness due to plastic relaxation.[3-5] Moreover, it is based on the selection of a specific substrate, whereas most applications require integration with silicon. Therefore, alternatives to the usual substrate engineering are needed, and some unconventional methods have been already developed.[6] For example, the inclusion of a secondary phase that segregates in nanocolumns in a ferroelectric BaTiO$_3$ (BTO) matrix and strains the latter can be used.[7] Without using secondary phases, residual stress, which is usual in BTO films deposited by energetic techniques like pulsed laser deposition or sputtering, can be used to modify BTO lattice parameters.[8-14] The resulting strain influences severely the ferroelectric properties and highly enhanced tetragonality (c/a > 1.1) and ferroelectric remnant polarization (> 50 μC/cm$^2$) has been reported in BTO films on SrTiO$_3$(001).[9] The control of strain and ferroelectric polarization by thin film deposition parameters would be a versatile strategy, alternative to the classic strain engineering, that could permit controlling strain and properties of ferroelectric films integrated with silicon.

It has been reported the dependence of lattice strain on laser fluency used to grow BTO films by pulsed laser deposition on $GdScO_3$(110).[8] The films showed differences in the Curie temperature although similar polarization, and it was proposed the existence of dipole defects in the films.[8] We present a different strategy of controlling the strain in BTO films, which generates a scalable switchable polarization with strain. The strategy is based on controlling the defects that generate strain by tuning the balance between thermodynamics and kinetics in the growth of BTO via modification of substrate deposition temperature. We focus on thin films integrated with Si(001),[15-17] and we show that epitaxial growth can be achieved in a broad deposition temperature window with a huge impact on the BTO tetragonality and ferroelectric polarization. In particular, high-quality *c*-oriented epitaxial BTO films are grown in a temperature window about 300 °C wide, permitting fine tuning of the *c*-axis (the polar axis) strain from 0% (high deposition temperature, favoring thermodynamics) to more than 2% (low deposition temperature, imposing kinetic limitations), and with the remnant polarization scaling with the c-axis. The growth strategy has permitted an unprecedented level of control of structural and ferroelectric properties of epitaxial BTO thin films on silicon, and also allows tailoring the strain in ferroelectric films on perovskite substrates.

**Results**

The specular XRD θ-2θ scans of the samples on Si(001) are shown in Fig. 1a. There are (00l) reflections from the Si substrate, YSZ, $CeO_2$ and $LaNiO_3$ (LNO) buffer layers, and BTO film, without peaks from spurious phases or other crystal orientations. It is remarkable that in spite of the broad range of BTO growth temperature $T_s$ (close to 400 °C, from $T_s$ = 375 °C to 750 °C), BTO is single (00l) oriented in all the samples. A zoomed region of the θ-2θ scans around the (002) reflections of BTO and LNO is in Fig. 1b. The solid and dashed vertical lines mark the position of the (002) and (200) reflections of bulk BTO, respectively. The intensity of the BTO(002) peak in the $T_s$ = 375 °C sample, the lowest BTO growth temperature, is much reduced in comparison with the other samples. To quantify the BTO crystallization dependence on deposition temperature, the BTO(002) peak intensity has been normalized to that of YSZ(002) peak of the corresponding sample. The dependence with $T_s$ of the intensity ratio, in logarithmic scale, is plotted in Fig. 2a. The intensity ratio increases sharply up to around 425 °C, with much lower dependence for higher temperatures. It indicates that the threshold temperature for BTO crystallization by pulsed laser deposition on the used buffer layers is around 375 °C. The dependences with $T_s$ (Fig. 2a, right axis) of the rocking curve (Δω, ω-scan) and particularly the width (Δ2θ, θ-2θ scan) of the BTO(002) reflection also reflect the onset temperature of crystallization. Moreover, the data do not indicate differences in crystal quality between the films deposited at temperature above 425 °C, with similar values of Δω (~1.1°) and Δ2θ (~0.4°).

The position of the BTO peak depends on $T_s$, being coincident with the bulk BTO(002) in the film grown at the highest temperature ($T_s$ = 750 °C) and at lower 2θ angles in the other samples. This result indicates that the BTO films are c-oriented, the *c*-axis strain depending on the substrate temperature. ϕ-scans and reciprocal space maps around asymmetrical reflections confirmed epitaxial growth. A sketch of the heterostructure is presented in Fig. 1c. Fig. 1d shows ϕ-scans around BTO(110), LNO(110), YSZ(220) and Si(220) reflections of the Ts = 675 °C sample. Each ϕ-scan shows a set of four peaks, the BTO and LNO peaks being shifted 45° with respect to the YSZ and Si ones. The lattice parameter of $CeO_2$ is almost coincident with Si, and the $CeO_2$(220) reflections overlap with the high intensity Si substrate peaks. It is concluded that the four layers are epitaxial, with cube-on-cube epitaxial relationship of $CeO_2$ and

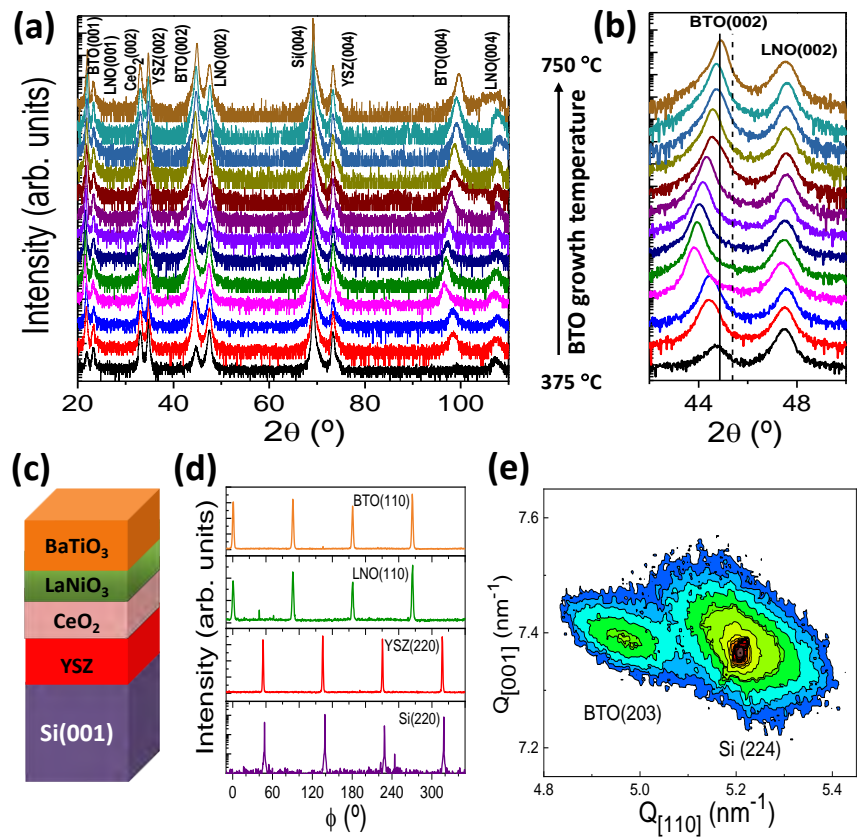

**Figure 1.** (a) XRD θ-2θ scans of the BTO/LNO/$CeO_2$/YSZ/Si(001) series. The intensity is plotted in logarithmic scale, and the diffractograms are shifted vertically for clarity, from the $T_s$ = 375 °C to the 700 °C sample. (b) Zoomed region of the θ-2θ scans around the (002) reflections of BTO and LNO. The vertical solid and dashed lines mark the position of the (002) and (200) reflections in bulk BTO, respectively. (c) Sketch of the heterostructure. (d) ϕ-scans around the BTO(110), LNO(110), YSZ(220) and Si(220) reflections of the $T_s$ = 675 °C sample. (e) Reciprocal space map around the BTO(203) and Si(224) reflections of the $T_s$ = 700 °C sample.

YSZ with Si(001), whereas BTO and LNO present an in-plane rotation of 45°.[15] Reciprocal space maps around the BTO(203) and Si(224) reflections of several samples were measured (the corresponding to the $T_s$ = 700 °C sample is in Fig. 1e). The BTO peak indicates homogeneous strain in all the measured samples (Supplementary Information Fig. S1). The in-plane lattice parameter is plotted (red triangles) against $T_s$ in Fig. 2b, together with the out-of-plane lattice parameter (blue squares, determined from the specular θ-2θ scans) of all BTO films on Si. The horizontal dashed lines indicate the values of the *a* (3.994 Å) and c (4.038 Å) parameters of bulk BTO. It is observed that the out-of-plane parameter increases first up to a strain $\varepsilon_{[001]}$ of 2.3% in the $T_s$ = 425 °C sample, and for higher deposition temperatures decreases monotonically reaching the bulk value of the c-axis in the $T_s$ = 750 °C sample. The strain displays minor variations if the films are either cooled down under high oxygen pressure (200 mbar instead of 0.2 mbar) or by additional annealing step (1 hour, 600 °C, 200 mbar). The strain and ferroelectric polarization data of these samples are presented in Supplementary Information S2. The effect of ex-situ annealing (1 hour, 200 mbar) on the lattice parameter $c$ of the $T_s$ = 450 °C and 600 °C samples was investigated too (Supplementary Information S3). The samples show almost negligible differences after annealing at 450 °C. The annealing at 600 °C has very small effect on the $T_s$ = 600 °C sample, and in the $T_s$ = 450 °C sample there is a small decrease of the out-of-plane parameter. Fig. S3c shows small dependence of the out-of-plane lattice parameter with the deposition temperature $T_s$, including the data corresponding to the *in-situ* and *ex-situ* annealing, proving the limited effects of annealing and signaling the high stability of the films.

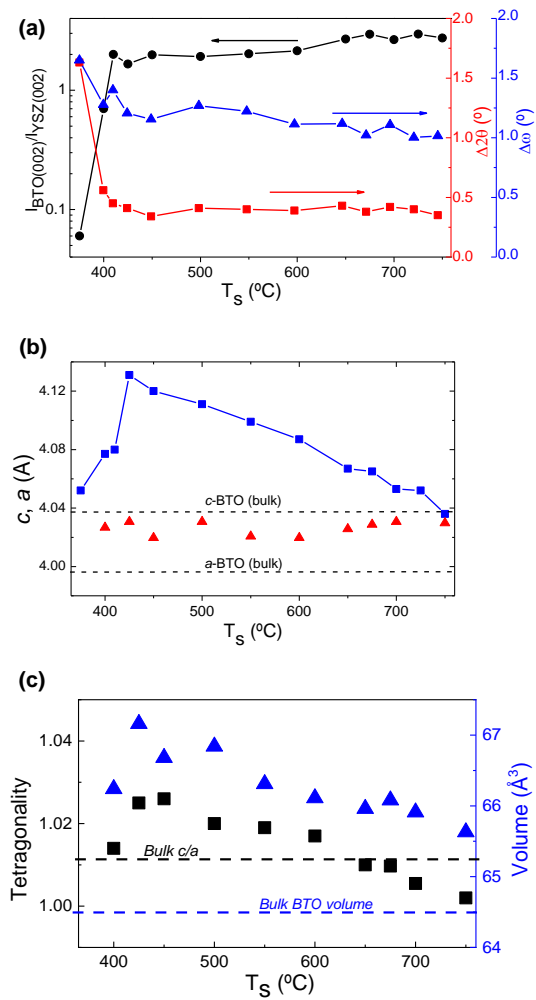

**Figure 2.** Dependence with the deposition temperature of BTO on LNO/CeO$_2$/YSZ/Si(001) of (a) intensity ratio between the BTO(002) and YSZ(002) peaks (black circles, left axis), and full width at half maximum of the 2θ- scan (red squares, right axis), and ω-scan (blue triangles, right axis) of the BTO(002) reflection. (b) Dependence of the out-of-plane lattice parameter (blue squares), and in-plane lattice parameter for selected samples (red triangles). The horizontal dashed lines indicate the *a*- and *c*-axes length in bulk BTO. (c) Unit cell tetragonality (left axis, black squares) and volume (right axis, blue triangles). The horizontal dashed lines indicate the c/a tetragonality and unit cell volume of bulk BTO.

The in-plane parameter of the BTO films deposited at different $T_s$ varies slightly between the samples, with values in the 4.021 – 4.032 Å range. Three factors can contribute to the expansion of the in-plane parameter: i) the epitaxial strain (it is not expected to be relevant due to the relatively large film thickness and lattice mismatch, anticipating a fast in-plane relaxation); ii) the point defects in the films (more relevant for the samples deposited at low $T_s$, although the unit cell is expected to expand mainly along the out-of-plane direction due to the epitaxial in-plane matching); and iii) the higher thermal expansion coefficients of BTO than of silicon, must induce a tensile stress when the films are cooled down to room temperature. The thermal mismatch stress is more important in the high $T_s$ samples. The combined effect of these factors results in similar in-plane parameter in the $T_s$ = 400 – 750 °C range. The unit cell tetragonality (c/a) and volume (Fig. 2c) increase with Ts up to 450 °C and 500°C, respectively, and for higher $T_s$ both decrease monotonically. The tetragonality is smaller than in bulk BTO in the samples deposited at $T_s$ higher than 650 °C. In contrast, the unit cell is expanded in all the samples, with high volume expansions ranging from around 1.9% (750 °C sample) to above 3.7 % (500 °C sample).

Topographic AFM images, 5 μm x 5 μm in size, of the $T_s$ = 375, 400, and 700 °C samples on Si are shown in Fig. 3a, 3b and 3c, respectively, with 1 μm x 1 μm images in the corresponding insets. The three images (and the corresponding ones to other films in the series) show islands with lateral size of a few tens of nm. The lateral size is around 40 nm in the $T_s$ = 375 °C and 700 °C samples, and around 80 nm in the $T_s$ = 400 °C sample. The morphology of the $T_s$ = 375 °C sample is very homogeneous, whereas in the $T_s$ = 700 °C sample there is agglomeration of a few islands of higher height in some areas. The morphology of the $T_s$ = 400 and the 410 °C samples (the AFM image of the latter not shown here) differs with respect to the other samples, with flat areas between the islands, which are notably larger in height than in the other samples. Notice in Fig. 3 (a-c) the differences in the z-scales. Indeed, the z-scale ranges in the topographic images are 7 nm in Fig. 3a and 3c, and 40 nm in Fig. 3b. The corresponding height profiles in Fig. 3e, 3f and 3g along the dashed lines shown in Fig. 3a, 3b, and 3c, evidence the huge difference in islands height. The root means square (rms) roughness of all the samples on Si, calculated from 1 μm x 1 μm areas, is plotted as a function of $T_s$ in Fig. 3d. There is a sharp increase from 0.56 nm to more than 5 nm as $T_s$ increases from 375 to 400 °C. The rms roughness, similarly high in the $T_s$ = 410 °C sample, decreases to around 1.6 nm and 0.92 nm in the $T_s$ = 425 and 450 °C samples, respectively, and is below 0.75 nm for the samples grown at higher temperatures. The peaky dependence of the roughness with deposition temperature is a signature of the onset of BTO crystallization above 375 °C, being the high roughness in the 400 - 425 °C samples likely due to the coexistence of well crystallized islands with other regions presenting lower crystal order or amorphous state. The (001) orientation in the $T_s$ = 400 °C sample, with in-plane epitaxial order confirmed by reciprocal space mapping (Supplementary Information Fig. S1) points to epitaxial columnar growth coexisting with not-epitaxial regions. On the other hand, the low surface roughness of the $T_s$ = 375 °C sample indicates that surface diffusivity at this temperature is low for epitaxial growth but high enough for surface smoothing.

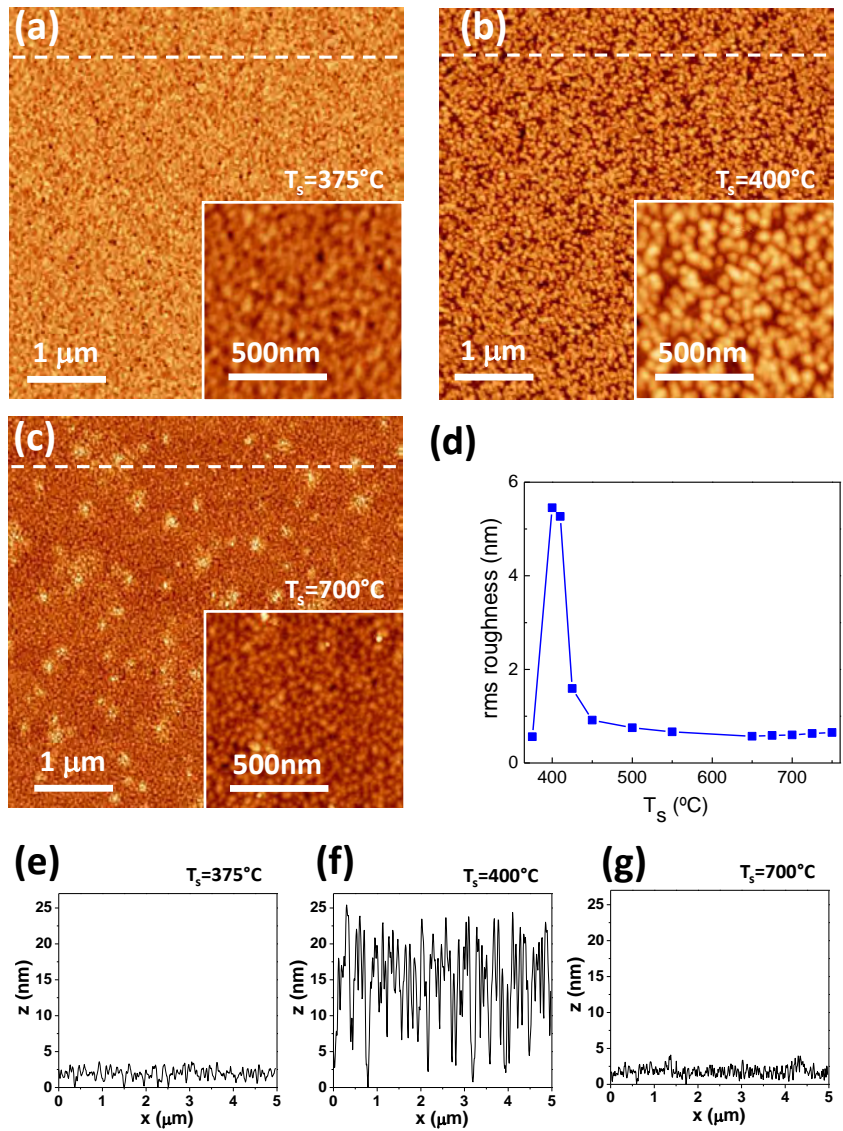

**Figure 3.** Topographic AFM images, 5 μm x 5 μm in size (inset: 1 μm x 1 μm) of the (a) $T_s$ = 375 °C (z-scale: 7 nm), (b) $T_s$ = 400 °C (z-scale: 40 nm), and (c) $T_s$ = 700 °C (z-scale: 7 nm) BTO films on LNO/CeO$_2$/YSZ/Si(001). Height profiles along the horizontal dashed lines shown in each 5 μm x 5 μm image are in (e), (f) and (g), respectively. (d) rms roughness as a function of $T_s$.

The BTO films deposited on Si at 410 °C or higher $T_s$ are ferroelectric. The ferroelectric polarization loops of a selection of the films are plotted in Fig. 4a. The loops were obtained from current - electric field measurement (the corresponding to the $T_s$ = 425 °C sample is presented in the inset). The remnant polarization $P_r$ and electrical coercive field $E_c$ of the films range within 3 - 11 $\mu C/cm^2$ and 70 – 160 kV/cm, respectively. The films deposited at the lowest temperatures, 375 and 400 °C, did not display current ferroelectric switching peaks. Fig. 4b presents the values of remnant polarization and electrical coercive fields of the BTO films on Si as a function of the deposition temperature. The threshold for ferroelectric behavior is $T_s$ = 410 °C, and for higher $T_s$ the remnant polarization increases sharply to a maximum value of around 11 $\mu C/cm^2$ ($T_s$ = 425 °C). For higher deposition temperatures, $P_r$ decreases monotonically with $T_s$. In the case of the coercive field the dependence is similar, with a reduction from 160 kV/cm ($T_s$ = 425 °C) to 70 kV/cm ($T_s$ = 750 °C).

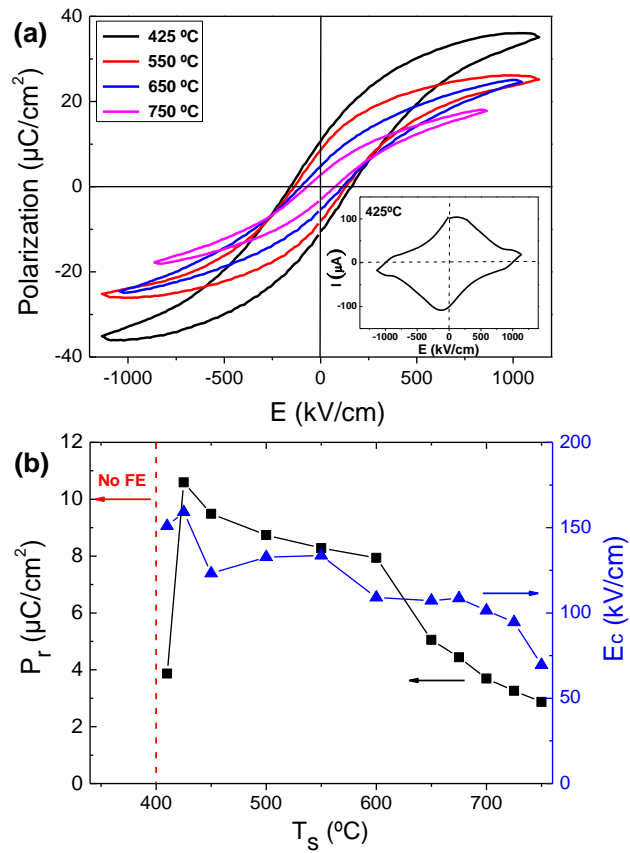

**Figure 4.** (a) Ferroelectric polarization loops of the $T_s$ = 425 °C, 550 °C, 650 °C and 750 °C BTO films on LNO/CeO$_2$/YSZ/Si(001), with the current (I) – electric field (E) curve corresponding to the $T_s$ = 425 °C sample in the inset. (b) Remnant polarization (black squares, left axis) and coercive field (blue triangles, right axis) as a function of $T_s$.

The leakage curves of the series of BTO films on Si are shown in Fig. 5a, and the values of leakage current at 45 and 225 kV/cm are plotted as a function of $T_s$ in Fig. 5b. The leakage current depends notably on $T_s$, particularly in the samples deposited at low temperatures. The $T_s$ = 375 °C sample was highly insulating and the small area of the contacts did not permit reliable measurements. The conductivity increases sharply with $T_s$, the $T_s$ = 450 °C sample being the most conductive of the series. In particular, the leakage current at 45 kV/cm increases from $10^{-7}$ A/cm$^2$ ($T_s$ = 400 °C) to around $3 \times 10^{-5}$ A/cm$^2$ ($T_s$ = 450 °C), and from $4 \times 10^{-7}$ A/cm$^2$ to $10^{-3}$ A/cm$^2$ at 225 kV/cm. Films deposited above $T_s$ = 450 °C are more insulating, with a monotonic reduction of the leakage current with Ts, presenting the $T_s$ = 750 °C samples leakage current below $1 \times 10^{-7}$ A/cm$^2$ and $4 \times 10^{-6}$ A/cm$^2$ at 45 and 225 kV/cm, respectively. We emphasize that the leakage of these films is comparable to state of the art BTO films on perovskite single crystalline substrates[5,7] or to thick epitaxial Pb(Zr,Ti)O$_3$ films on Si(001).[18]

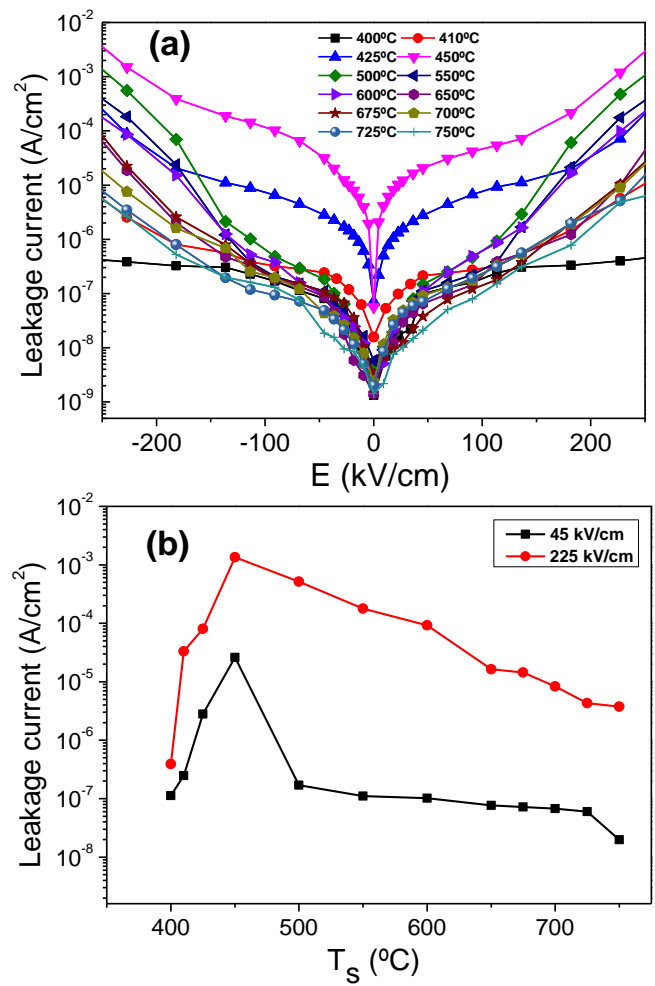

**Figure 5.** (a) Leakage current curves of the BTO films on LNO/CeO$_2$/YSZ/Si(001). (b) Leakage current at 45 kV/cm (black squares) and 225 kV/cm (red circles) as a function of the deposition temperature.

The impact of the deposition temperature on the properties of BTO is schematized in Fig. 6. Above the crystallization threshold at 375 °C there is a temperature window around 75 °C wide, up to $T_s \sim 450$ °C where films have a large strain and ferroelectric polarization, but are rough and present high leakage. XRD and AFM characterization point to in-homogeneous crystallinity as the cause. From $T_s \sim 450$ °C to $\sim 750$ °C, the films are very flat and are highly insulating. In this temperature window, about 300 °C wide, the BTO films are c-oriented and show a monotonic variation of the *c*-axis parameter and the ferroelectric polarization. The unit cell of BTO is highly expanded, likely due to the presence of point defects.[19-22] The influence of the deposition temperature on these defects permits the control of the *c*-axis parameter and the ferroelectric polarization.

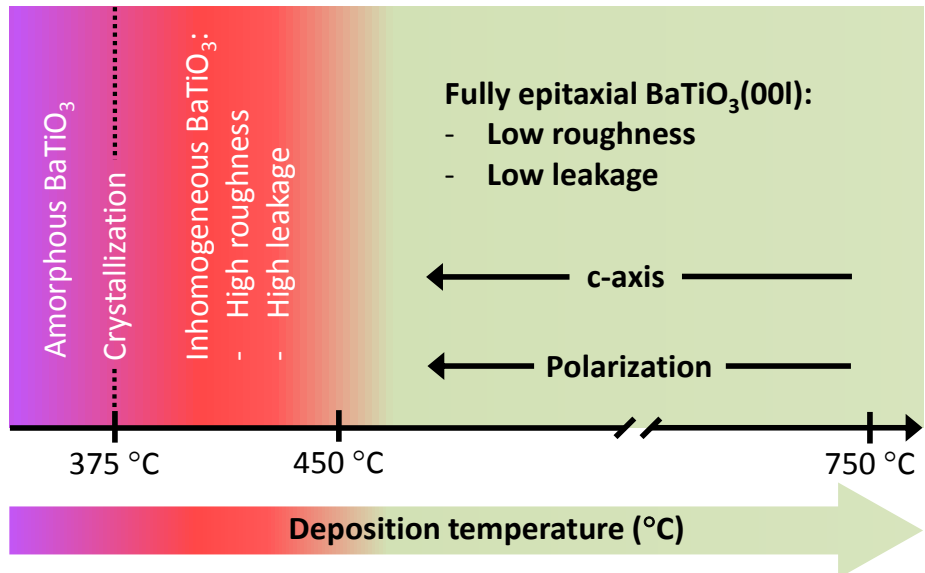

**Figure 6.** Schematics of the influence of the deposition temperature on the crystallinity and properties of $BaTiO_3$ thin films on $LNO/CeO_2/YSZ/Si(001)$.

In contrast to extended defects, the identification of randomly distributed point defects in oxide thin films by transmission electron microscopy is challenging and indirect methods are typically used.[21] Here we have used the photocurrent induced by illumination with 405 nm photons (energy of 3.06 eV, close but smaller than the optical gap of BTO as a probe for point defect concentration. Indeed, it is expected that point defects shall induce in-gap states promoting larger optical absorption of sub-bandgap photons and enlarge the photocurrent. The photocurrent increases with $T_s$ from $8 \times 10^{-3}$ $\mu A/cm^2$ ($T_s = 400$ °C sample) up to above 0.2 $\mu A/cm^2$ in the $T_s = 425$ °C and $T_s = 450$ °C samples, and then decreases progressively to around 0.1 $\mu A/cm^2$ in the $T_s = 700$ °C and $T_s = 750$ °C samples (Fig. 7). The out-of-plane lattice parameter and the photocurrent display a very similar dependence on the deposition temperature. The photocurrent is higher in the more strained films (Fig. 7, inset), supporting that cell expansion is caused by the defects responsible of the photoresponse. On the other hand, other authors reported shifted (imprinted) ferroelectric polarization loops in epitaxial BTO films grown by pulsed laser deposition on perovskite substrate and suggested that the internal electric field to be related to the presence of aligned defect dipoles.[8] To discern the presence of defect dipoles in our BTO films we prepared symmetric

LNO/BTO/LNO capacitors on (001)-oriented LaAlO$_3$ (LAO) at T$_s$ = 700 °C. Capacitors with symmetric electrodes were grown on purpose to minimize imprint due differences on screening properties and work functions of electrodes. The ferroelectric polarization loop (Supplementary Information Fig. S4) shows an imprint field (≈50 kV/cm), pointing towards the top LNO electrode. The presence of an internal electric field in symmetric capacitors is consistent with the existence of aligned defect dipoles in our BTO films.

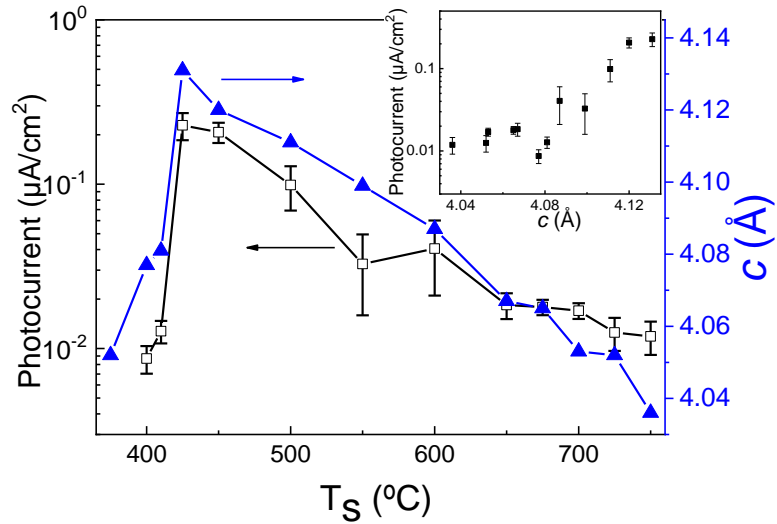

**Figure 7.** Dependence with the deposition temperature of the photocurrent (left axis, open black squares) and the out-of-plane lattice parameter (right axis, solid blue triangles). Photocurrent was measured in at least ten pairs of contacts in each film. The vertical lines in the photocurrent data indicate the ranges of values measured. Inset: photocurrent against out-of-plane lattice parameter.

It is remarkable that the control of strain and polarization by the deposition temperature has been demonstrated in thick BTO films (thicker than 100 nm) integrated epitaxially with Si(001). This is clearly the most convenient substrate for applications, and the demonstration that integration of epitaxial BTO with high polarization can deposited at temperature as low as 450 °C is also of relevance towards the integration in silicon chips. Moreover, the growth strategy here presented should also be valid for BTO films grown on single crystalline oxide substrates, which are commonly used as platforms to grow BTO films. To confirm this statement we have deposited two BTO/LNO bilayers on LAO at 550 and 675 °C. Due to the moderately small lattice mismatch (~1.3%) between the LNO electrode and the LAO substrate, the XRD specular θ-2θ scans of the bilayers (Fig. 8a) only display (00l) reflections. The zoom of the scans around the (002) reflections shows narrow BTO peaks at positions that confirm the expected c-orientation and expanded $c$-axis, being the strain in the T$_s$ = 550 °C sample (c = 4.123 Å) higher than in the T$_s$ = 675 °C one (c = 4.096 Å). Reciprocal space maps around asymmetrical (103) reflections are shown in Supplementary Information Fig. S5. The LNO bottom electrode is fully strained, whereas BTO presents in-plane lattice parameter coincident with the a-axis of bulk BTO. It indicates that the films are plastically relaxed, and that the defects cause anisotropic unit-cell deformation, with expansion of the unit cell along the out-of-plane direction. Topographic AFM images show that the films are very flat, and in the case of the $T_s$ = 675 °C sample rms roughness below 0.4 nm and morphology of terraces and steps can be observed in the 5 µm x 5 µm image (Fig. 8c) in spite of the very small islands (Fig.

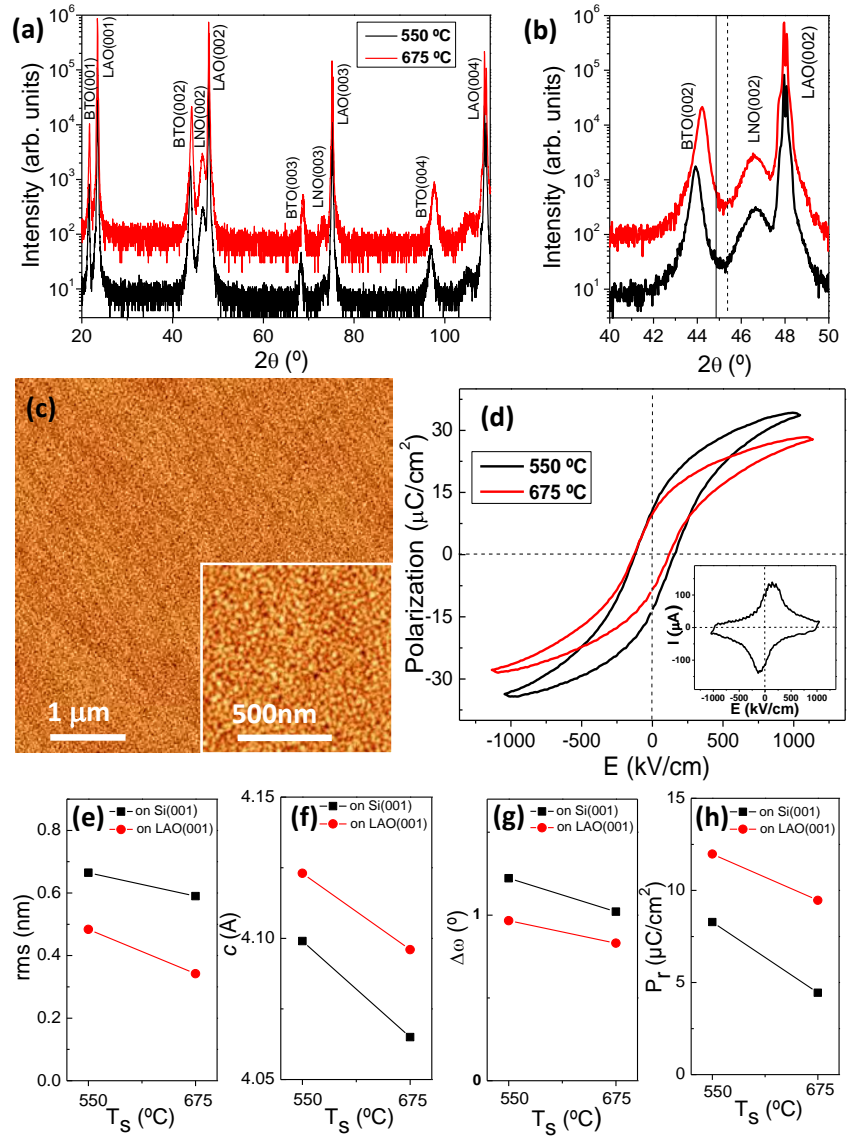

**Figure 8.** (a) XRD θ-2θ scans of the films deposited on LNO/LaAlO$_3$(001) at $T_s$ = 550 °C (black line) and $T_s$ = 675 °C (red line). The intensity scale is logarithmic and the diffractograms are shifted vertically for clarity. (b) Zoomed region of the θ-2θ scans around the (002) reflections of BTO, LNO and LAO. The vertical solid and dashed lines mark the position of the BTO(002) and BTO(200) reflections in bulk BTO, respectively. (c) Topographic AFM 5 µm x 5 µm image (inset: 1 µm x 1 µm) of the film deposited at $T_s$ = 675 °C (z-scale: 3 nm). (d) Ferroelectric polarization loops of the $T_s$ = 550 °C (black line) and 675 °C films on LAO, with the current – electric field curve corresponding to the 425 °C sample in the inset. Comparison of rms roughness (e), out-of-plane lattice parameter (f), full width at half maximum of the ω-scan (g), and remnant polarization (h) of BTO films on Si (black squares) and LAO (red circles) at $T_s$ = 550 °C and $T_s$ = 675 °C.

8c, inset), usually present in BTO films.[15,23,24] The ferroelectric loops (Fig. 7d) show larger polarization in the $T_s$ = 550 °C film, in agreement with its higher lattice strain. The values of rms roughness, c-axis parameter, BTO(002) rocking curve width and remnant polarization are plotted (red circles) against $T_s$ (550 °C or 675 °C) in Fig. 8e, 8f, 8g, and 8h, respectively. They include the data of the corresponding samples on Si (black squares). Obviously, the BTO films deposited on Si(001) and LAO(001) substrates display the same trend of film morphology (rms), crystallinity (*c*-axis parameter and rocking curve) and polarization.

**Discussion**

In summary, *c*-oriented BTO films have been epitaxially integrated with Si(001) with tailored strain of the polar axis in a broad range from 0 to above 2%. In contrast to the usual strain engineering methodology, the selection of a particular strain is achieved without necessity of selecting a specific substrate, and it is not limited to very thin ferroelectric films. To accomplish this, using high energy pulsed laser deposition technique, the substrate temperature is used as a knob within a wide temperature window of about 300 °C to grow epitaxial BTO films with the desired c-axis strain. The ferroelectric polarization scales with the strain, and, concomitantly, it is determined by the selected deposition temperature. Using this methodology, either on Si(001) or on perovkite oxide single crystals, epitaxial BTO films with large polarization and very small roughness and electrical leakage are produced. Of technological relevance, it is demonstrated that these properties are achieved integrating BTO with Si(001) at temperatures as low as 450 °C.

**Methods**

**Thin films deposition**. BTO films were deposited on Si(001) substrates using a multilayer buffer. The first buffer layer, yttria-stabilized zirconia (YSZ), was grown on Si(001) without removing the native silicon oxide, and $CeO_2$, LNO and BTO were sequentially deposited. The heterostructures were fabricated in a single process by pulsed laser deposition (KrF excimer laser). The buffer layers were grown using the same parameters, including substrate temperature of 800 °C for YSZ and CeO2, and 700 °C for LNO, and oxygen pressure of 4x10-4 mbar for YSZ and $CeO_2$, and 0.15 mbar for LNO; additional experimental conditions are reported elsewhere.[15,17,25] The deposition temperature was measured using a thermocouple inserted in the middle of the heater block. The thicknesses of the BTO, LNO, $CeO_2$ and YSZ layers are 110, 30, 20, and 60 nm, respectively. BTO films were deposited under a dynamic oxygen pressure of 0.02 mbar, being the laser frequency of 5 Hz. A series of 13 samples was prepared varying the substrate temperature in steps of 50 °C between 400 and 750 °C, and additional samples were deposited at $T_s$ = 375, 410, 425, 675 and 725 °C. Moreover, two BTO/LNO bilayers were prepared on LAO (001) at 550 and 675 °C. After deposition, the BTO films were cooled down to room temperature under an oxygen atmosphere of 0.2 mbar. Two additional films deposited on Si(001) at $T_s$ = 700 °C were cooled down under 200 mbar of oxygen, adding for one of an in-situ at 600 °C for 1 hour. The effect of ex-situ annealing (1 hour, 200 mbar) was investigated on the $T_s$ = 450 °C and 600 °C films on Si(001). Both samples were annealed sequentially two times at 450 °C and 600 °C for 1 hour under 200 mbar.

**Structural and electrical characterization**. The crystal structure was characterized by X-ray diffraction (XRD), determining the out-of-plane lattice parameter of BTO from symmetrical θ-2θ scans (Cu Kα radiation), and measuring rocking curves of the BTO(002) reflection. Reciprocal space maps around BTO(203) and Si(224) of selected samples were measured by high resolution XRD using Cu Kα1 radiation to obtain the in-plane lattice parameter of BTO, and ϕ-scans around BTO(110), LNO(110), YSZ(220) and Si(220) were performed to determine the epitaxial relationships. The surface morphology of BTO was characterized by atomic force microscopy (AFM) in dynamic mode. Platinum top electrodes, 20 nm thick and 60 μm x 60 μm in size, were deposited using dc magnetron sputtering through stencil masks. Ferroelectric polarization loops and leakage current were measured at room temperature in top-top configuration[26] (two BTO capacitors were measured in series, contacting two top Pt electrodes and using the conducting LNO buffer layer as common bottom electrode) by means of an AixACCT TFAnalyser2000 platform. Ferroelectrics loops were obtained by sweeping an electric field at a constant rate with a frequency of 10 kHz and measuring the current, using the dielectric leakage current compensation (DLCC) to minimize leakage current effects.[27] Leakage current curves were measured using the TFAnalyser2000 platform (or an electrometer for the samples grown at $T_s$ = 410 °C or below, which are highly insulating), using 3 s integration time, and averaging I-V curves increasing and decreasing the voltage. Short-circuit photocurrent was measured by illuminating the sample with blue laser of wavelength 405 nm (Shenzhen 91 Laser Co.). Photoinduced current was monitored as a function of time (t) switching on the illumination at 5 s and off at 15 s. The spot diameter was of 200 μm safely illuminating the measured electrode.

**References**

1. Choi, K. J. *et al.* Enhancement of ferroelectricity in strained BaTiO₃ thin films. *Science* **306**, 1005-1009 (2004).

2. Schlom, D.G. *et al.* Strain tuning of ferroelectric thin films. *Annu. Rev. Mater. Res.* **37**, 589-626 (2007).

3. Sun, H. P., Tian, W., Pan, X. Q., Haeni, J. H. & Schlom, D. G. Evolution of dislocation arrays in epitaxial BaTiO$_3$ thin films grown on (100) SrTiO$_3$. *Appl. Phys. Lett.* **84**, 3298-3300 (2004).

4. Kawai, M. *et al.* Critical thickness control by deposition rate for epitaxial BaTiO3 thin films grown on SrTiO$_3$(001). *Appl. Phys. Lett*. **102**, 114311 (2007).

5. Chen, Y. B. *et al.* Interface structure and strain relaxation in BaTiO$_3$ thin films grown on GdScO$_3$ and DyScO$_3$ substrates with buried coherent SrRuO$_3$ layer. *Appl. Phys. Lett.* **91**, 252906 (2007).

6. Damodaran, A.R. *et al.* New modalities of strain-control of ferroelectric thin films. *J. Phys. Condens. Matter.* **28**, 1 (2016).

7. Harrington, S. A. *et al.* Thick lead-free ferroelectric films with high Curie temperatures through nanocomposite-induced strain. *Nature Nanotechnol.* **6**, 491-495 (2011).

8. Damodaran, A. R., Breckenfeld, E., Chen, Z., Lee, S. & Martin, L. W. Enhancement of Ferroelectric Curie Temperature in BaTiO$_3$ Films via Strain-Induced Defect Dipole Alignment. *Adv. Mater.* **26**, 6341-6347 (2014).

9. Fina, I. *et al.* Direct magnetoelectric effect in ferroelectric - ferromagnetic epitaxial heterostructures. *Nanoscale* **5**, 8037-8044 (2013).

10. Sinsheimer, J. *et al.* In-situ x-ray diffraction study of the growth of highly strained epitaxial BaTiO$_3$ thin films. *Appl. Phys. Lett.* **103**, 242904 (2013).

11. Yanase, N., Abe, K., Fukushima, N. & Kawakubo, T. Thickness dependence of ferroelectricity in heteroepitaxial BaTiO$_3$ thin film capacitors. *Jpn. J. Appl. Phys.* **38**, 5305-5308 (1999).

12. Petraru, A., Pertsev, N. A., Kohlstedt, H., Poppe, U. & Waser, R. Polarization and lattice strains in epitaxial BaTiO$_3$ films grown by high-pressure sputtering. *J. Appl. Phys.* **101**, 114106 (2007).

13. Fu, D. *et al.* High-T$_c$ BaTiO$_3$ ferroelectric films with frozen negative pressure states. arXiv: 1102.4473v1.

14. Khestanova, E. *et al.* Untangling electrostatic and strain effects on the polarization of ferroelectric superlattices. *Adv. Funct. Mater.* **26**, 6446-6453 (2016).

15. Scigaj, M. *et al.* Ultra-flat BaTiO$_3$ epitaxial films on Si(001) with large out-of-plane polarization. *Appl. Phys. Lett.* **102**, 112905 (2013).

16. Scigaj, M. *et al.* High ferroelectric polarization in c-oriented BaTiO$_3$ epitaxial thin films on SrTiO$_3$/Si(001). *Appl. Phys. Lett.* **109**, 122903 (2016).

17. Scigaj, M. *et al.* Monolithic integration of room-temperature multifunctional $BaTiO_3$-$CoFe_2O_4$ epitaxial heterostructures on Si(001). *Sci. Reports* **6**, 31870 (2016).

18. Dekkers, M. *et al.* Ferroelectric properties of epitaxial Pb(Zr,Ti)$O_3$ thin films on silicon by control of crystal orientation. *Appl. Phys. Lett.* **95**, 012902 (2009).

19. Hennings, D. F. K., Metzmacher, C. & Schreinemacher, B. S. Defect chemistry and microstructure of hydrothermal barium titanate. *J. Am. Ceram. Soc.* **84**, 179-182 (2001).

20. Tsur, Y. & Randall, C. A. Point defect concentrations in barium titanate revisited. *J. Am. Ceram. Soc.* **84**, 2147-2149 (2001).

21. Tuller, H. L.; Bishop, S. R. Point defects in oxides: tailoring materials through defect engineering. *Annu. Rev. Mater. Res.* **41**, 369-398 (2011).

22. Pramanick, A., Prewitt, A. D., Forrester, J. S. & Jones, J. L. Domains, domain walls and defects in perovskite ferroelectric oxides: a review of present understanding and recent contributions. *Crit. Rev. Sol. Sta. Mater. Sci.* **37**, 243-275 (2012).

23. Chen, X. *et al.* Ultrathin $BaTiO_3$ templates for multiferroic nanostructures. *New J. Phys.* **13**, 083037 (2011).

24. Dix, N. *et al.* Large out-of-plane ferroelectric polarization in flat epitaxial $BaTiO_3$ on $CoFe_2O_4$ heterostructures. *Appl. Phys. Lett.* **102**, 172907 (2013).

25. de Coux, P. *et al.* Mechanisms of epitaxy and defects at the interface in ultrathin YSZ films on Si(001). *CrystEngComm* **4**, 7851-7855 (2012).

26. Liu, F., Fina, I., Bertacco, R. & Fontcuberta, J. Unravelling and controlling hidden imprint fields in ferroelectric capacitors. Sci. Re*p.* **6**, 25028 (2016).

27. Fina, I. *et al.* Non-ferroelectric contributions to the hysteresis cycles in manganite thin films: a comparative study of measurement techniques. *J. Appl. Phys.* **109**, 074105 (2011).

**Acknowledgements**

Financial support from the Spanish Ministry of Economy and Competitiveness, through the "Severo Ochoa" Programme for Centres of Excellence in R&D (SEV-2015-0496) and the MAT2017-85232-R, MAT2014-56063-C2-1-R, and MAT2015-73839-JIN projects, and from Generalitat de Catalunya (2014 SGR 734) is acknowledged. IF acknowledges Juan de la Cierva – Incorporación postdoctoral fellowship (IJCI-2014-19102) from the Spanish Ministry of Economy and Competitiveness. JL is financially supported by China Scholarship Council (CSC) with No. 201506080019. JL work has been done as a part of his Ph.D. program in Materials Science at Universitat Autònoma de Barcelona.

**Author Contributions**

FS designed and supervised the experiments. RS deposited the thin films. Structural analysis by XRD and AFM was done by JL. Electric characterization was done by JL with support of IF. FS wrote the manuscript with collaboration of JF and IF. All authors discussed the data and commented on the paper.

## Additional Information

**Supplementary Information** is available.

**Competing financial interests**: The authors declare no competing financial interests.

**Tailoring Lattice Strain and Ferroelectric Polarization of Epitaxial BaTiO$_3$ Thin Films on Si(001)**

Jike Lyu, Ignasi Fina, Raúl Solanas, Josep Fontcuberta & Florencio Sánchez

Institut de Ciència de Materials de Barcelona (ICMAB-CSIC), Campus UAB, Bellaterra 08193, Barcelona, Spain

**Supplementary Information**

**Supplementary Information S1: XRD reciprocal space maps of BTO films on Si(001)**

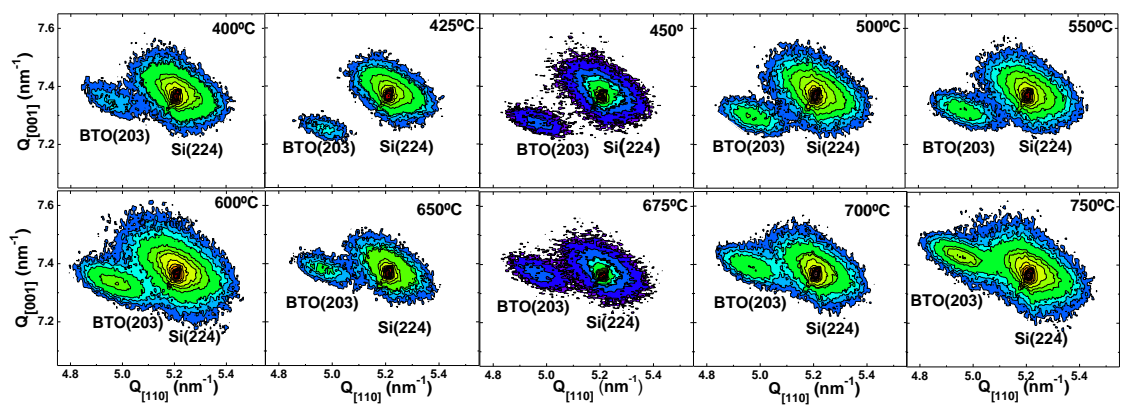

**Figure S1:** XRD reciprocal space maps (RSM) of BTO films on Si(001) deposited at various substrate temperature $T_s$ (indicated in the top right side of each panel). The mapped reciprocal space contains asymmetrical BTO(203) and Si(224) reflections. The RSM we recorded using Cu K$\alpha_1$ using same acquisition conditions(except for the $T_s$ = 450 °C and 675 °C samples, recorded with shorter acquisition time).

**Supplementary Information S2:**

**Influence on strain and polarization of high oxygen pressure during cooling down and of in-situ annealing**

The films reported in the manuscript were cooled down to room temperature under the same conditions. An oxygen pressure of 0.2 mbar was introduced immediately after the growth, and the substrate heater power was switched off. Two additional BTO/LNO/CeO$_2$/YSZ/Si(001) samples, deposited at $T_s$ = 700 °C, were cooled down under a different process after the growth. Sample (1): the oxygen pressure introduced at the end of the growth was 200 mbar, and the film was cooled down to room temperature under this pressure and with the substrate heater power switched off. Sample (2): the oxygen pressure introduced at the end of the growth was 200 mbar, and it was cooled to room temperature after a dwell time of 1 hour at 600 °C. The two different cooling down processes do not cause great effects on the BTO strain and polarization (Figure S2) respect to the reference sample cooled down under 0.2 mbar of oxygen (out-of-plane lattice parameter c = 4.053 Å, remnant polarization $P_r$ = 3.7 μC/cm$^2$. In the case of Sample 1 (cooled down under 200 mbar), c and $P_r$ are 4.049 Å and 3.1 μC/cm$^2$, respectively, and in sample 2 (cooled down under 200 mbar, and in-situ annealing at 600 °C for 1 hour) c and $P_r$ are 4.043 Å and 3.3 μC/cm$^2$, respectively. Figures S2a and S2b show, respectively, the $T_s$ dependence of the c-axis values and $P_r$ values, and the corresponding values for samples (1) and (2) are plotted (blue squares). The $P_r$ values of the series is plotted against c in Figure S2c (data corresponding to samples (1) and (2) are plotted as blue squares). There is a clear linear dependence, and samples (1) and (2) scale well in the series, suggesting a slight reduction in the amount of defects in the two samples.

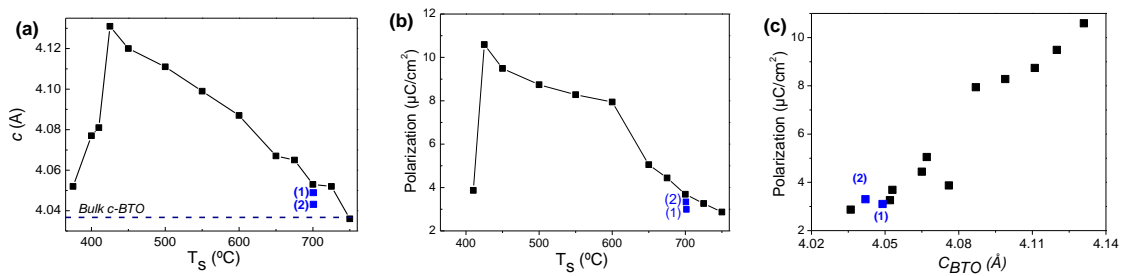

**Figure S2**: Dependence with the deposition temperature $T_s$ of BTO on LNO/CeO$_2$/YSZ/Si(001) of (a) out-of-plane lattice parameter c, and (b) remnant polarization. (c) Remnant polarization plotted against $T_s$. Black squares correspond to the samples discussed in the manuscript (they were cooled down to room temperature under an oxygen pressure of 0.2 mbar introduced immediately after the growth). Samples labelled 1 and 2 (blue squares) were cooled down under a different procedure. Sample (1): the oxygen pressure introduced at the end of the growth was 200 mbar, and it was cooled down to room temperature under this pressure. Sample (2): the oxygen pressure introduced at the end of the growth was 200 mbar, and it was cooled to room temperature under this pressure after a dwell time of 1 hour at 600 °C.

**Supplementary Information S3:**

**Influence on strain of ex-situ annealing**

The $T_s$ = 450 °C and 600 °C films on Si(001) were cut in two pieces, and only one of them was used for two sequential annealings. were annealed during 1 hour under 200 mbar of atomic oxygen. The cut piece of each sample was measured by XRD (θ-2θ scan), annealed at 450 °C during 1 hour under 200 mbar of atomic oxygen, measured by XRD, annealed at 600 °C during 1 hour under 200 mbar of atomic oxygen, and finally measured by XRD.

Figure S3 shows the θ-2θ scans (a). The samples show almost negligible differences after annealing at 450 °C. The annealing at 600 °C has very small effect on the $T_s$ = 600 °C sample, and in the $T_s$ = 450 °C sample there is a small decrease of the out-of-plane parameter. The variation of the lattice parameters with the annealings is plotted in (b). (c) Dependence of the out-of-plane lattice parameter with the deposition temperature $T_s$, including the data corresponding to the *ex-situ* annealed samples (open triangles) and the *in-situ* annealed samples (open squares).

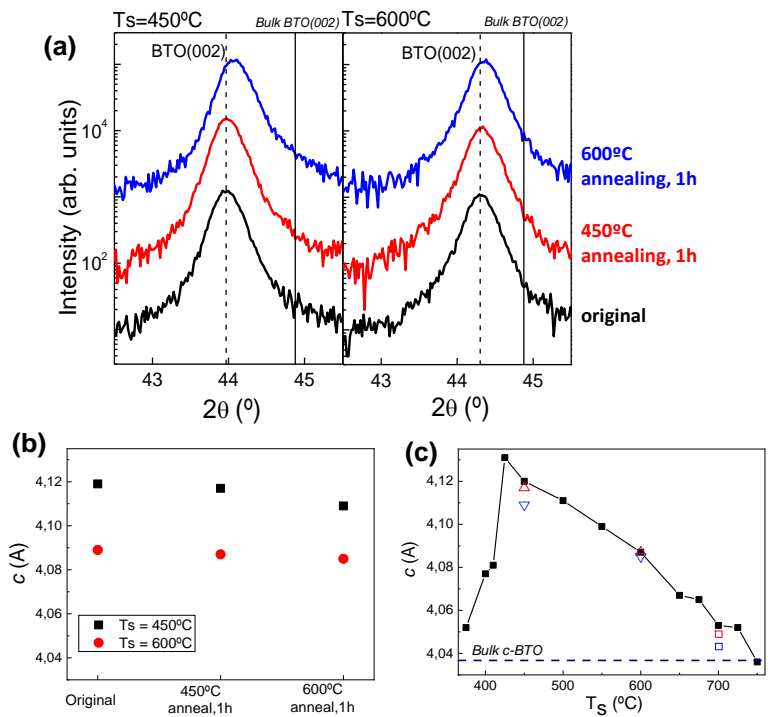

**Figure S3**: XRD θ-2θ scan of the (a) $T_s$ = 450 °C (left panel) and 600 °C (right panel) films on Si(001) as-deposited (black curves), after annealing at 450 °C (red curves), and after 600 °C (blue curves). The patterns are shifted vertically for clarity. The vertical solid line mark the position of the BTO(002) reflection in bulk BTO, and the vertical dashed line mark the corresponding position of the BTO(002) reflections of the as-deposited films. (b) Out-of-plane lattice parameter c of the as-deposited and annealed films. (c) Dependence of the out-of-plane lattice parameter c with the deposition temperature $T_s$ of BTO on LNO/CeO$_2$/YSZ/Si(001), including the values of the $T_s$ = 450 °C and $T_s$ = 600 °C samples after the *ex-situ* first annealing at 450 °C (open red triangles up) and second annealing at 600 °C (open blue triangles down). The corresponding data of the *in-situ* annealed $T_s$ = 700 °C samples are plotted (open red and blue squares).

**Supplementary Information S4: Polarization loops of symmetric on LaNiO$_3$/BaTiO$_3$/LaNiO$_3$ capacitors on LaAlO$_3$(001)**

J. Lyu, I. Fina, R. Solanas, J. Fontcuberta, and F. Sánchez , Scientific Reports 8, 495 (2018)

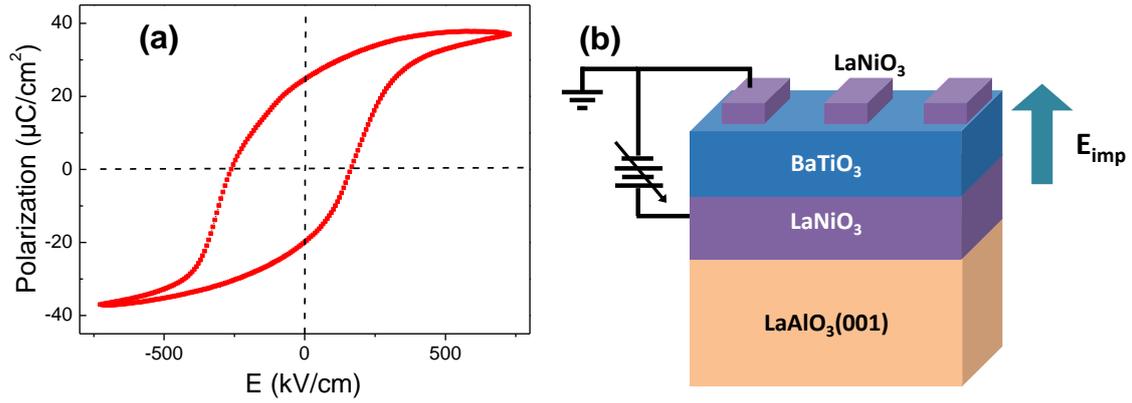

**Figure S4**: (a) Polarization ferroelectric loop (DLCC measurement, 1 kHz, circular top electrode of diameter 200 μm) of a symmetric $LaNiO_3/BaTiO_3/LaNiO_3$ capacitor on $LaAlO_3(001)$. A sketch is plotted in (b). The BTO film was deposited at 700 °C. The top LNO electrode was deposited by pulsed laser deposition at 700 °C and 0.15 mbar of oxygen, using a steel shadow mask in contact with the BTO film. The ferroelectric loop was measured applying positive voltage to bottom LNO electrode and grounding a top LNO electrode. The observed horizontal negative shift corresponds to an imprint field of ≈50 kV/cm, which is directed from the bottom electrode towards the top electrode.

**Supplementary Information S5: XRD reciprocal space maps of BTO films on $LaAlO_3(001)$**

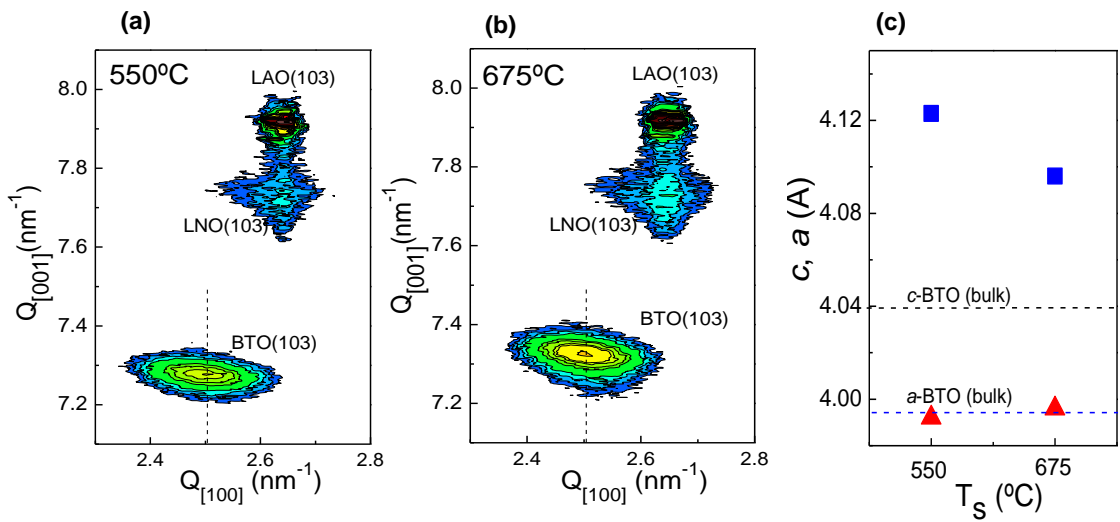

**Figure S5:** XRD reciprocal space maps (RSM) of BTO films deposited on $LNO/LaAlO_3(001)$ at $T_s$ = 550 °C (a) and $T_s$ = 675 °C (b). The mapped reciprocal space contains asymmetrical (103) reflections of BTO, LNO and LAO. The RSMs we recorded using Cu $K\alpha_1$. The vertical dashed line indicates the $Q_{[100]}$ coordinate for bulk BTO (a-axis). (c) Out-of-plane (blue squares) and in-plane (red triangles) lattice parameters of the two BTO films. The horizontal dashed lines indicate the corresponding parameters for bulk BTO.

19